\documentclass[a4paper,10pt]{article}
\usepackage[utf8]{inputenc}

\usepackage[T1]{fontenc}
\usepackage[utf8]{inputenc}

\usepackage{authblk}

\usepackage{tensor}
\usepackage{amsmath}
\usepackage{amsfonts}
\usepackage{amssymb}
\usepackage{amsbsy}
\usepackage{bm}
\usepackage{graphicx}

\usepackage[centering,
marginparwidth=0in,
textwidth=6.5in,
marginparsep=2em,
top=2.5cm,
bottom=2.5cm,
]{geometry}

\usepackage{tensor}
\usepackage{tikz-cd}
\usepackage{stmaryrd}

\newcommand{\eg}{e.\,g.}%
\newcommand{\wrt}{w.r.t.}%

\newcommand{\surf}{\mathcal{S}}

\newcommand{\TangentBundle}[2]
{\mathrm{T}_{#1}{#2}}
\newcommand{\TensorBundle}[3][1={n},2={}]
{\mathrm{T}^{#1}_{#2}{#3}}

\newcommand{\normalCmp}{\nu}
\newcommand{\normal}{\boldsymbol{\nu}}

\newcommand{\wb}{\boldsymbol{w}}%
\newcommand{\bb}{\boldsymbol{b}}%
\newcommand{\eb}{\boldsymbol{e}}%
\newcommand{\xb}{\boldsymbol{x}}%
\newcommand{\ub}{\boldsymbol{u}}%
\newcommand{\Ub}{\boldsymbol{U}}%

\newcommand*{\rom}[1]{\textup{\uppercase\expandafter{\romannumeral#1}}}

\newcommand{\shapeOperator}{\boldsymbol{B}}%
\newcommand{\meanCurvature}{\mathcal{H}}%
\newcommand{\Div}{\operatorname{div}}

\newcommand{\trace}{\operatorname{tr}}%
\newcommand{\Id}{\boldsymbol{I}}

\newcommand{\ProjSurf}{\boldsymbol{P}}

\newcommand{\laplaceSurf}{\triangle_{\surf}}

\newcommand{\GradSurf}{\nabla_{\surf}}

\newcommand{\DivP}{\Div_{\ProjSurf}}
\newcommand{\GradP}{\nabla_{\ProjSurf}}
\newcommand{\LaplaceBeltrami}{\laplaceSurf}

\newcommand{\DivC}{\Div_{C}}
\newcommand{\GradC}{\nabla_{C}}

\newcommand{\surfaceParam}{\mathbf{X}}
\newcommand{\surfaceParamCmp}{X}

\newcommand{\energy}{\mathcal{F}}

\newcommand{\stressC}{\sigma}
\newcommand{\stress}{\boldsymbol{\stressC}}

\newcommand{\density}{\rho}

\newcommand{\velocityCmp}{u}
\newcommand{\velocity}{\ub}
\newcommand{\relMaterialVelocity}{\wb}

\newcommand{\pressure}{p}

\newcommand{\angleVelocity}{\omega}

\newcommand{\inertiaTensor}{\mathbf{\mathbb{I}}}
\newcommand{\inertiaTensorCmp}{\mathbb{I}}

\newcommand{\viscosity}{\eta}

\newcommand{\bendingRigidity}{\kappa}

\newcommand{\ReynoldsNumber}{\mathrm{Re}}
\newcommand{\BendingNumber}{\mathrm{Be}}

\renewcommand{\Re}{\ReynoldsNumber}

\begin{document}

\title{Stability of rotating equilibrium states of fluid deformable surfaces}

\author[1]{Michael Nestler}
\author[1,2,3]{Axel Voigt}

\affil[1]{Institute of Scientific Computing, TU Dresden, Dresden, 01062, Germany}
\affil[2]{Center for Systems Biology Dresden (CSBD), Pfotenhauerstra{\ss}e 108, Dresden, 01307, Germany}
\affil[3]{Cluster of Excellence Physics of Life (PoL), Dresden, 01062, Germany}


\maketitle  
\begin{abstract}
We consider rotating equilibrium states of fluid deformable surfaces. These states are characterized by a force balance between centrifugal and bending forces, involve surface Killing vector fields and are independent on the surface viscosity. Considering a continuum description based on the incompressible surface Navier–Stokes equations with bending forces and conserved enclosed volume we numerically demonstrate that these rotating equilibrium states can be reached, but also that these states are not stable. Any perturbation in shape or rotating flow field leads to dissipation and destroys the rotating equilibrium states. After breaking symmetry the evolution reaches other rotating states with a lower energy for which the symmetry axis and the rotation axis are not aligned. Such flow fields could be characterized by three-dimensional Killing vector fields. However, also these states are not stable. Based on these numerical results we postulate a cascading mechanism of ’disturbance - force balance reconfiguration - dissipation’ that contains various rotating equilibrium states as transient configurations but eventually leads to the classical equilibrium shapes of the Helfrich energy.
\end{abstract}

\section{Introduction}

Fluid deformable surfaces are soft materials exhibiting a solid–fluid duality: while they store elastic energy when
stretched or bent, like solid shells, under in-plane shear they flow as two-dimensional, viscous fluids. With this solid–fluid duality any shape change contributes to tangential flow and vice versa any tangential flow on a curved surface induces shape deformations. Such surfaces play an essential role in biology, see e.g. \cite{mayer2010anisotropies,mietke2019self,salbreux2017mechanics,Voigt_JFM_2019}. The mathematical description of fluid deformable surfaces has been introduced in \cite{Arroyoetal_PRE_2009,salbreux2017mechanics,torres2019modelling,reuther2020numerical} and the highly nonlinear system of partial differential equations is numerically solved in \cite{torres2019modelling,reuther2020numerical,krause2023numerical,Krause_PAMM_2023}.

In \cite{reuther2020numerical,krause2023numerical} the existence of rotating symmetric equilibrium states is postulated for fluid deformable surfaces. These states are independent on the surface viscosity and are further elaborated in \cite{olshanskii2023equilibrium} where under the assumption of rotational symmetry and a balance of centrifugal and bending forces a wide range of shapes and associated Killing fields are proposed. These shapes differ from classical equilibrium shapes without flow \cite{seifert1997configurations}. Here we ask if these states can be reached using the full model without enforcing any symmetry and if these states are stable. We will demonstrate by numerical experiments and theoretical arguments that the shapes suggested in \cite{olshanskii2023equilibrium} can be reached providing rotational symmetric initial states, but that these states are unstable. Small perturbations in shape or flow field lead to dissipation and destroys these states. The rotational symmetry gets broken and the system evolves towards new configurations. These states are again characterized by rotation. However, the axis of rotation and the symmetry axis of the surface no longer coexist. These equilibrium states have a lower energy than than the once proposed in \cite{olshanskii2023equilibrium}. However, with the same argument as before, we propose that also these states are not stable and small perturbations in shape or flow field will lead to further reconfiguration and finally convergence towards the classical equilibrium shape without flow \cite{seifert1997configurations}. But even if rotating equilibrium states are not stable they can be transient states in the evolution of fluid deformable surfaces.

The paper is structured as follows. We introduce the notation and the surface operators and explain the model of fluid deformable surfaces in section \ref{sec:methods}. Section \ref{sec:experiments} provides information on the numeric experiments and results on the evolution and stability of potential rotating equilibrium states are discuss in section \ref{sec:discussion}. Concluding remarks are presented in section \ref{sec:conclusion}.

\section{Mathematical model and numerical methods}
\label{sec:methods}

We consider $\surf(t) \subset \mathbb{R}^3$ as a two dimensional embedded surface described along a parametrization $\surfaceParam(t)$. The surface is endowed with an outward pointing normal $\normal$ and geometric properties like shape operator $\shapeOperator  =- \GradC \normal$, mean curvature $\meanCurvature = \trace \shapeOperator$ and surface projection $\ProjSurf = \Id - \normal\normal$. The material surface velocity is denoted by $\velocity: \, \surf \rightarrow \mathbb{R}^3$, including tangential and normal components. For such non-tangential vector fields $\velocity$ we use the G{\"u}nther derivative $\GradP$, the tangential derivative defined along the gradient $\nabla$ of the embedding space by $\GradP \velocity = \ProjSurf \left[\nabla \velocity^e\right] \ProjSurf$ with associated divergence $\DivP \velocity = \trace \left[\GradP \velocity^e\right]$, where $\velocity^e$ refers to an extension of $\velocity$ to $\mathbb{R}^3$ with $\velocity^e$ constant in normal direction. This notion of derivatives can be related to the covariant derivative $\GradSurf$ for tangential vector fields, see \cite{nitschke2012finite,brandner2022finite}. Furthermore, we use the componentwise surface gradient $\GradC$ with $\GradC \stress = [\nabla \stress^e] \ProjSurf$ and $\DivC(\stress \ProjSurf) = \trace(\stress \ProjSurf)$. For scalar fields, as the surface pressure $\pressure$, the tangential derivative coincides with the covariant derivative $\GradSurf \pressure = \GradP \pressure = \ProjSurf [\nabla \pressure^e]$.

The density $\density$, bending rigidity $\bendingRigidity$, viscosity $\viscosity$, characteristic velocity $\Ub$ and characteristic length $L$ allow to define the Reynolds number $\ReynoldsNumber= \density \Ub L / \viscosity$ and the bending capillary number $\BendingNumber = \density \Ub^2 L^2 / \bendingRigidity$. With these numbers the non-dimensional total free energy $\energy$, consisting of a kinetic energy $\energy_{kin}$ and a Helfrich energy $\energy_{bend}$ reads
\begin{align}
\energy = \energy_{kin} + \energy_{bend} = \frac{1}{2} \int_{\surf} \velocity^2 \, \mathrm{d}\surf + \frac{1}{2 \BendingNumber}\int_{\surf} \meanCurvature^2 \, \mathrm{d}\surf \label{eq:freeEnergy}
\end{align}
with spontaneous curvature $\meanCurvature_0 = 0$ and neglected Gaussian curvature terms.
Mass and momentum balance lead to the desired fluid deformable surface model, which combines the incompressible surface Navier-Stokes equations, see \eg\ \cite{arroyo2009relaxation,nitschke2012finite,Kobaetal_QAM_2017,jankuhn2018,nitschke2019hydrodynamic} with classical Willmore flow \cite{Dziuk_NM_1990}. As usual the surface pressure $\pressure$ can be interpreted as a Lagrange multiplier to enforce mass conservation. By $\DivP \velocity = 0$ we enforce local inextensibility and therefore surface area conservation. We further restrict the evolution by preserving the enclosed volume. This is done by adding another Lagrange multiplier $\gamma$ to enforce $\int_{\surf} \velocity \cdot \normal \,\mathrm{d}\surf=0$, see \cite{torres2019modelling,krause2023numerical} for details. The full model for fluid-deformable surfaces reads
\begin{align}
\begin{aligned}
\partial_t \velocity + \nabla_{\relMaterialVelocity}\velocity &= -\GradSurf p-p\meanCurvature\normal + \frac{2}{\Re}\DivC \stress(\velocity)-\gamma\normal + \mathbf{b} \\
        \DivP \velocity &= 0\\
        \int_{\surf} \velocity\cdot\normal \, \mathrm{d}\surf &= 0 
\end{aligned} \label{eq:momentummassbalance}
\end{align}
with the rate-of-deformation tensor $\stress(\velocity) = 1/ 2 (\GradP \velocity + [\GradP \velocity]^T)$ and the convection term $[\nabla_{\relMaterialVelocity}\velocity]_i = \GradSurf \velocityCmp_i \cdot  \relMaterialVelocity$ \wrt\ the relative material velocity $\relMaterialVelocity = \velocity - \partial_t \surfaceParam$, which follows from a mesh regularizing tangential flow 
\begin{align}
\begin{aligned}
\partial_t \surfaceParam \cdot \normal & = \velocity \cdot \normal \\
\meanCurvature \normalCmp_i & = \LaplaceBeltrami \surfaceParamCmp_i,     
\end{aligned} \label{eq:surfaceevolution}
\end{align}
where $\LaplaceBeltrami$ denotes the Laplace Beltrami operator, see \cite{barrett2008parametric,krause2023numerical} for details. Following \cite{arroyo2009relaxation,reuther2016incompressible}, the bending forces resulting from $\energy_{bend}$ reads
\begin{align}
    \bb = \frac{1}{\BendingNumber}\left( -\LaplaceBeltrami \meanCurvature - \meanCurvature(\| \shapeOperator\|^2 - \frac{1}{2}\trace \shapeOperator) \right) \normal . \label{eq:elasticforces}
 \end{align}
The numerical approach to solve eqs. \eqref{eq:momentummassbalance} - \eqref{eq:elasticforces} is described in detail in \cite{krause2023numerical}. It is based on the surface finite element method \cite{dziuk2013finite} extended to vector-valued surface partial differential equations \cite{nestler2019finite}.  Implementation is done by the finite element framework AMDiS \cite{vey2007amdis,witkowski2015software} base on DUNE \cite{SanderDune2020,bastian2021dune} using the grid manager AluGrid \cite{dunealugrid:16} and the CurvedGrid library \cite{praetorius2020dunecurvedgrid}. All linear systems are solved by a direct solver using the external library PETSc. We will explore this approach to investigate the equilibrium states proposed in \cite{olshanskii2023equilibrium}. 

\section{Computational experiments to explore rotating equilibrium states}
\label{sec:experiments}

\subsection{Problem setup}
Under the assumption of rotational symmetry three conditions are identified for equilibrium states in \cite{olshanskii2023equilibrium}. They read 
$$\velocity \cdot \normal = 0, \quad  \stress(\velocity) = 0, \quad p - \frac{1}{2} |\velocity|^2 = p_0,$$
with same constant $p_0$ and are used to derive a shape equation and an equation for the angular velocity. Besides the trivial states with $\velocity = 0$ these conditions also allow for equilibrium states with $\velocity \neq 0$. Associating the surface pressure $p$ with the surface tension, the last condition can be interpreted as a flow dependent surface tension. The first two conditions define so-called 'tangentially rigid' motions, which correspond to surface Killing vector fields \cite{Nitschkeetal_TFI_2017,Pruessetal_JEE_2021}. We can interpret such Killing vector fields as rigid body rotations, \wrt\ to a symmetry axis of $\surf$. They induce centrifugal forces which balance the bending forces leading to equilibrium shapes unlike the ones described in \cite{seifert1997configurations}.  

From the perspective of the full fluid deformable surface model in eqs. \eqref{eq:momentummassbalance} - \eqref{eq:elasticforces} without any symmetry assumptions several questions arise: Are the proposed equilibrium states in \cite{olshanskii2023equilibrium} reachable? Are these rotational symmetric states stable in the sense that small perturbations in shape or flow field decay and a rotational symmetric shape recovers with symmetry and rotation axis aligned? And, do other states exist with $\velocity \neq 0$ which are energetically more favorable? We discuss these questions along two experiments considering the relaxation of rotational symmetric initial conditions. We therefore consider a prolate and an oblate spherical ellipsoid endowed with an initial rigid body rotation $\omega_0$, \wrt\ z-axis, where body symmetry and rotational axis align. Furthermore, we choose a set of parameters, compatible with \cite{olshanskii2023equilibrium}, with $\ReynoldsNumber=1,\, \BendingNumber=1$. Furthermore, we consider a reduced volume \cite{seifert1997configurations} of $\hat{V}=0.75$, for which dumbbell and biconcave shapes are stable shapes for $\omega=0$, see \cite{seifert1997configurations}, yet the dumbbell shape poses a global minimum of the bending energy. The initial geometry for the prolate is set by the major axes $[0.61, 0.61,2.02]$ and the oblate is defined by $[1.27,1.27,0.47]$. We apply an initial angular velocity of $\omega_0=4$. As numerical approach we consider the one proposed in \cite{krause2023numerical}. We use the same numerical parameters, only the the time step size differs and is set to $\tau=0.00025$. One critical issue is the considered mesh regularization in \eqref{eq:surfaceevolution}, successfully applied in \cite{Dziuk_NM_2008,Baenschetal_JCP_2005,Hausseretal_JSC_2007,Krause_PAMM}. It enables evaluation of rotating meshes only for a limited time. To avoid numerical issues, we restrict our evaluation to the time-span where monotony in $\energy$ is ensured. This is sufficient to answer most of the above questions.

To asses symmetry properties of the computed shapes $\surf$ and to quantify similarities to results in \cite{olshanskii2023equilibrium} we define a set of auxiliary quantities. To obtain an estimate on the angular velocity $\angleVelocity$ \wrt\ z-axis we evaluate
\begin{align}
    \eb_{\angleVelocity} & = \eb_z \times \eb_r = \frac{1}{r} [-y,x,0], \quad r = \sqrt{x^2 + y^2}, \quad 
    \angleVelocity(t) = \frac{1}{|\surf|}\int_{\surf} \frac{1}{r} \velocity(t) \cdot \eb_{\angleVelocity} \, \mathrm{d}\surf,
\end{align}
with $\mathbf{x} = [x,y,z]$. Furthermore, to asses the rotational symmetry of shapes $\surf$ we evaluate the inertia tensor $\inertiaTensor$ and its eigenvalues $\lambda_i$. Along the standard definition of the inertia tensor, see \eg\ \cite{chernousko2017evolution}, and $\xb_c = [x_c,y_c,z_c]$ the center of mass we evaluate
\begin{align}
\begin{aligned}
    \inertiaTensorCmp_{xx} & = \int_{\surf} \left( (y-y_c)^2 + (z-z_c)^2 \right) \,\mathrm{d}\surf,\qquad \inertiaTensorCmp_{yy} = \int_{\surf} \left( (x-x_c)^2 + (z-z_c)^2 \right) \,\mathrm{d}\surf, \\
    \inertiaTensorCmp_{zz} & = \int_{\surf} \left( (x-x_c)^2 + (y-y_c)^2 \right) \,\mathrm{d}\surf, \qquad 
    \inertiaTensorCmp_{yx} =\inertiaTensorCmp_{xy} = - \int_{\surf}  (x-x_c)(y-y_c) \,\mathrm{d}\surf, \\ 
    \inertiaTensorCmp_{xz} & = \inertiaTensorCmp_{zx} = - \int_{\surf}  (x-x_c)(z-z_c) \,\mathrm{d}\surf, \qquad
    \inertiaTensorCmp_{zy} = \inertiaTensorCmp_{yz} = - \int_{\surf}  (y-y_c)(z-z_c) \,\mathrm{d}\surf .
\end{aligned}
    \label{eq:inertiaTensorDef}
\end{align}
The eigenvector associated with the largest eigenvalue can be used to identify an axis of elongation in the shape and a necessary condition for a rotational symmetric shape is the existence of two identical eigenvalues of $\inertiaTensor$. The symmetry axes thus corresponds to the eigenvector of the remaining eigenvalue. To obtain a numerically feasible condition we consider the eigenvalue pair $(\lambda_a, \lambda_b)$ with minimum distance $\min_{ I,J, I\neq J}|\lambda_I -\lambda_J|$ and define a measure for the deviation from symmetry as
\begin{align}
e(\lambda) = \frac{|\lambda_a - \lambda_b|}{\langle \lambda \rangle}, \quad \langle \lambda \rangle = \sum_{i=1}^3 \frac{\lambda_i}{3} \label{eq:relSymmetryDeviation}
\end{align}
with $e(\lambda)=0$ for exact symmetry and $e(\lambda)=1$ for strong deviation from rotational symmetry. For $e(\lambda) < 1e-4$ we will consider a surface as numerically symmetric and with $e(\lambda) < 1e-2$ as numerically close to symmetric. We compare these data with the the rotational symmetric surface $\surf_O$ and the corresponding angular velocity $\angleVelocity$ proposed in \cite{olshanskii2023equilibrium}. The shapes are compared along a slice of the surface \wrt\ the symmetry axis (z-axis) and slice normal to it (x-axis). We parameterize both resulting curves $l_O(\surf^O)$ and $l_E(\surf)$ by polar coordinates $(r,\phi)$ and define the $L^2$ distance between the curves by 
\begin{align}
    E(L^2) = \left(\int_0^{2\pi} |r_O(\phi) - r_E(\phi)|^2 \,\mathrm{d}\phi \right)^{1/2}. \label{eq:ltwoDistToOlshanski}
\end{align}
Obviously this definition is only meaningful as long as $\surf$ is close to a rotational symmetric shape.


\subsection{Numerical results}

\begin{figure}[!h!]
    \centering
    \includegraphics[width=0.9\linewidth]{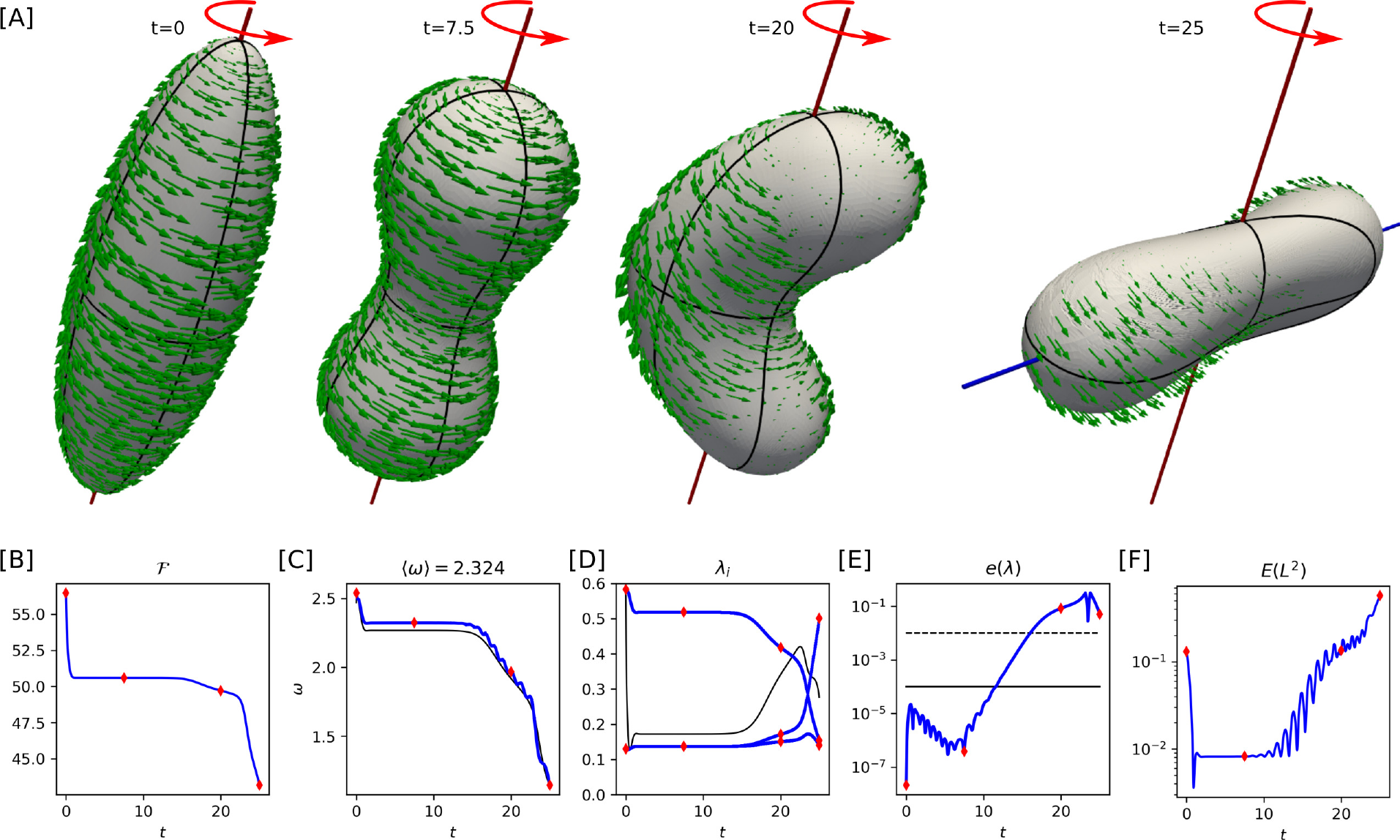}
    \caption{\textbf{Relaxation of a Prolate Ellipsoid $[0.61, 0.61,2.02]$:} [A] Characteristic snapshots of surface $\surf$ (black lines indicate slices \wrt\ principal axes of $\mathbb{R}^3$) and velocity $\velocity$ (encoded in scaled green arrows), snapshots corresponds to $t=0,\, 7.5,\, 20,\, 25$ (from left to right). Red arrows indicate axis and direction of rotation. Blue lines indicate axis of elongation, where at $t=0,\,7.5,\, 20$ elongation and rotation axis are aligned. [B] Transient of free energy $\energy$ as defined in eq. \eqref{eq:freeEnergy}. Red diamonds correspond to snapshot times shown in [A], same apply to plots [C]-[F]. [C] estimated angular velocity \wrt\ z-axis (blue line), $\langle \omega \rangle$ is obtained by averaging estimated values for $t\in[5,10]$. For comparison the scaled transient of $\energy_{kin}$ (black line) is shown. [D] Eigenvalues of inertia tensor $\inertiaTensor$ as defined in eq. \eqref{eq:inertiaTensorDef}. For comparison the scaled transient of $\energy_{bend}$ (black line) is shown. [E] Transient of relative deviation from rotational symmetry $e(\lambda)$ as defined in eq. \eqref{eq:relSymmetryDeviation}. Black line and dashed line indicate thresholds for symmetric and close to symmetry shapes. The thresholds are crossed the first time at $t_{1e-4}=11.6$ and $t_{1e-2}=16.0$. [F] Evolution of $L^2$ distance to corresponding shape proposed in \cite{olshanskii2023equilibrium}.}  
    \label{fig:prolate}
\end{figure}

For the initial prolate geometry, see Figure \ref{fig:prolate}-[A] we observe an evolution in three distinct phases. For $t \in[0,2.5]$ a fast relaxation towards the predicted dumbbell shape  \cite{olshanskii2023equilibrium} for $\langle \angleVelocity \rangle = 2.324$ occurs, see Figure \ref{fig:prolate}-[A]$t=7.5$. Within this relaxation phase the minimum of the similarity measure $E(L^2)$ is reached, see Figure \ref{fig:prolate}-[F]. This first phase includes a steep decay in kinetic and bending energy, see Figure \ref{fig:prolate}-[B] and a convergence of the angular velocity, see Figure \ref{fig:prolate}-[C]. Rotational symmetry is preserved in this phase, Figure \ref{fig:prolate}-[D,E] for the constant eigenvalues and the measure for deviation from symmetry, respectively. The evolution is followed by a second phase for $t\in[2.5, 16]$ which is characterized by a quasi constant free energy, a constant angular velocity and constant eigenvalues, see Figure \ref{fig:prolate}-[B,C,D], respectively. Yet, the logarithmic plots for measures $e(\lambda)$ and $E(L^2)$ indicate for $t>10$ an increasing loss of symmetry and increasing distance to the predicted equilibrium shape for $\langle \angleVelocity \rangle = 2.324$. After continuous increase of disturbances in symmetry, at $t > 16$ the evolution enters phase three, characterized by accelerating decay of free energy and strong shape changes, see Figure \ref{fig:prolate}-[A] $t=20,25$. In this phase, parts of the surface move radially outwards, \wrt\ rotational axis, while the surface is compressed in direction of the rotation axis. This shape transformation is associated with a strong increase of bending energy, which is compensated by a decay in kinetic energy and a decay in the approximated angular velocity $\angleVelocity$, see Figure \ref{fig:prolate}-[B,C], respectively. At the late stage of this transformation an elongated shape emerges, which is characterized by new dominant eigenvector/eigenvalue pairs. The elongation direction is now perpendicular to the rotation axis. This elongated shapes rotate quasi rigid \wrt\ the unchanged rotation axis. At the end of the evaluation time domain $t\in [0,25]$ the free energy is still decaying at a significant rate, such that the observed transformation has not reached its final configuration.

\begin{figure}[!ht]
    \centering
    \includegraphics[width=0.9\linewidth]{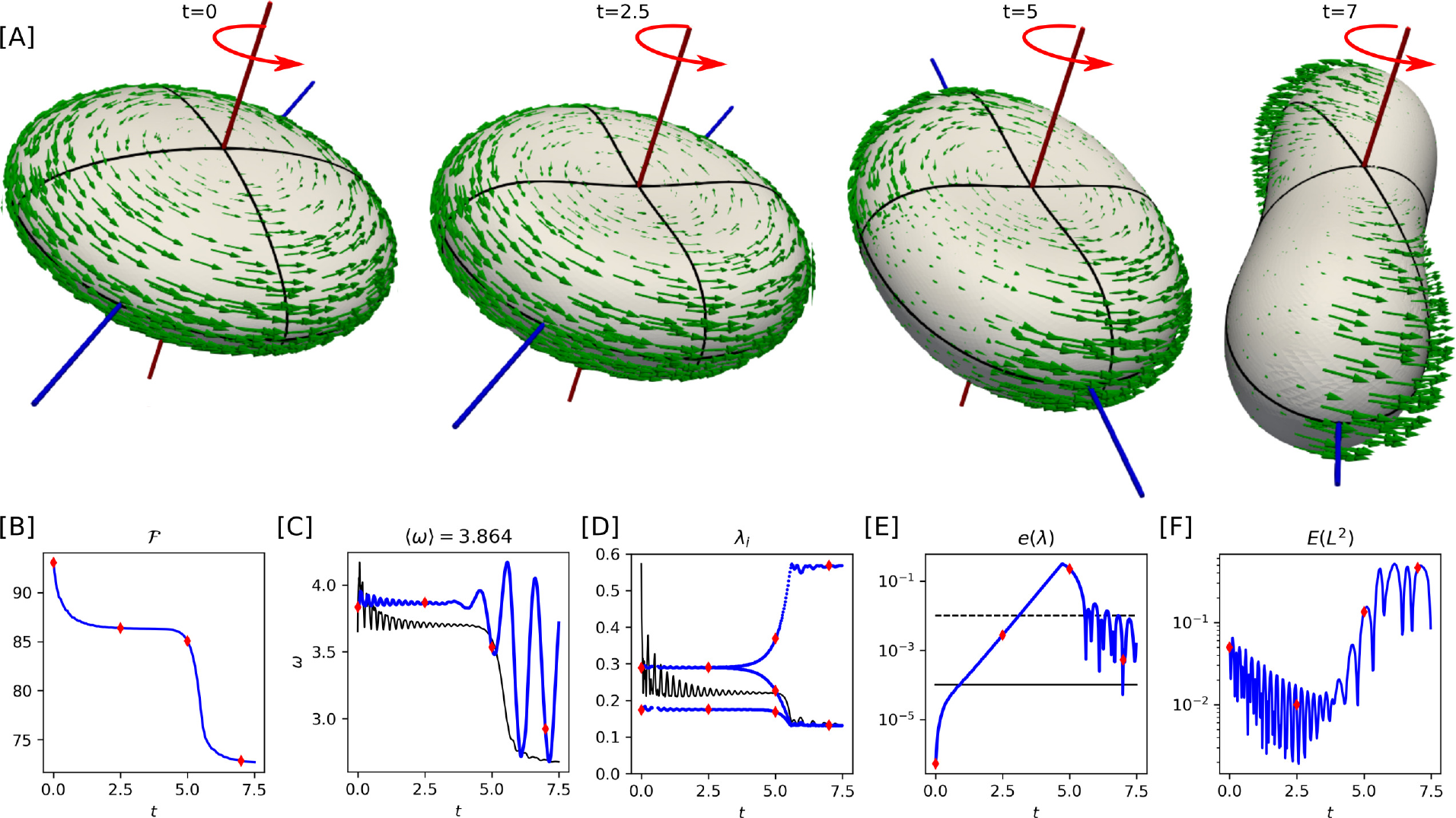}
    \caption{\textbf{Relaxation of an Oblate Ellipsoid $[1.27,1.27,0.47]$:} [A] Characteristic snapshots of surface $\surf$ (black lines indicate slices \wrt\ principal axes of $\mathbb{R}^3$) and velocity $\velocity$ (encoded in scaled green arrows), snapshots corresponds to $t=0,\, 2.5,\, 5,\, 7.0$ (from left to right). Red arrows indicate axis and direction of rotation. Blue lines indicate axis of elongation. At $t=0,\,2.5$ the elongation axis is indefinite in plane perpendicular to rotation axis due to roational symmetry. [B] Transient of free energy as defined in eq. \eqref{eq:freeEnergy}. Red diamonds correspond to snapshot times shown in [A], same apply to plots [C]-[F]. [C] estimated angular velocity \wrt\ z-axis, $\langle \omega \rangle$ is obtained by averaging estimated values for $t\in[2,3]$. For comparison the scaled transient of $\energy_{kin}$ (black line) is shown. [D] Eigenvalues of inertia tensor $\inertiaTensor$ as defined in eq. \eqref{eq:inertiaTensorDef}. For comparison the scaled transient of $\energy_{bend}$ (black line) is shown. [E] Transient of relative deviation from rotational symmetry $e(\lambda)$ as defined in eq. \eqref{eq:relSymmetryDeviation}. Black line and dashed line indicate thresholds for symmetric and close to symmetry shapes. The thresholds are crossed the first time at $t_{1e-4}=0.9$ and $t_{1e-2}=3.1$. [F] Evolution of $L^2$ distance to corresponding shape proposed in \cite{olshanskii2023equilibrium}.}  
    \label{fig:oblate}
\end{figure}

We consider the same experiment for the oblate geometry, see Figure \ref{fig:oblate}-[A], with similar results. As for the prolate geometry three distinct phases of relaxation can be identified. The first phase is again characterized by maintaining the rotational symmetry and relaxation towards the predicted  biconcave shape in \cite{olshanskii2023equilibrium} for $\langle \angleVelocity \rangle = 3.864$, see Figure \ref{fig:oblate}-[B,C,D,E,F]. Yet, contrary to the prolate geometry, we observe fluctuations between biconcave shapes with varying center intrusion depth, also visible in the oscillations of kinetic and bending energy as well as in estimated angular velocity in Figure \ref{fig:oblate}-[B,C], respectively. Due to the slow decay of fluctuations, we observe a short plateau $t\in[2.5,3.1]$ with quasi stationary free energy, shape, pure 'tangentially rigid' velocity and limited deviations for the predicted shape. Similar to the prolate experiment the continuous increasing loss of symmetry indicates the transition to the third phase. For $t \in [3.1,5.5]$, we observe the breakdown of rotational symmetry where parts of the shape are moved radial outwards, \wrt\ rotational axis, while remaining parts are moved radial inwards and an elongated shape, perpendicular to the rotational axis, forms, see Figure \ref{fig:oblate}-[A] at $t=5$. Contrary to the prolate geometry only a small compression of the shape in direction of the rotation axis is observed, see Figure \ref{fig:oblate}-[D]. Furthermore, we observe the stabilization of the elongated shape for $t>5.5$, with decaying shape changes, see also limited variations of eigenvalues in Figure \ref{fig:oblate}-[D], and decreasing rate of dissipation in $\energy$ in Figure \ref{fig:oblate}-[B]. 

\subsection{Discussion}
\label{sec:discussion}
We have performed a set of numerical experiments, for the parameter set of $\hat{V}=0.75$ and $\angleVelocity_0=4$, to test the predictions in \cite{olshanskii2023equilibrium} in a setting without enforcing rotational symmetric shapes. According to these experiments, the rotational symmetric shapes predicted in \cite{olshanskii2023equilibrium} can indeed be reached dynamically starting from initial values with rotational symmetry. This has been demonstrated along with two quantitative measures $e(\lambda)$ and $E(L^2)$. After a fast relaxation the full model converges to the rotational symmetric states, where symmetry axis and rotation axis are aligned, and a balance of centrifugal and bending forces for a tangential rigid velocity field is obtained. Yet, the mechanism of balancing centrifugal and bending forces is also the main driver in breaking the rotational symmetry, assumed in \cite{olshanskii2023equilibrium}. We observed the force balance to be very sensitive to local disturbances as any imbalance is self reinforcing. As an example, consider a surface element placed slightly more radial outward than the rotational symmetric shape would prescribe. At this element the force balance is disturbed with a resulting force pushing the element even further radial outward. Due to volume conservation, such local and radial outward movement has to be compensated by an inward motion of another surface element. Therefore the shape change extends and the small disturbance in symmetry is not recovered. This starts a cascade of self reinforcing shape changes and promotes the elongation perpendicular to the rotation axis. This reinforcing shape changes are only limited by bending energy. We postulate such self reinforcing mechanism as general, such that we expect shapes elongated perpendicular to the rotating axis are always present in the relaxation of rotating equilibrium states where rotational symmetry can not be guaranteed.

This allows to conclude that rigid body rotations of fluid deformable surfaces are not stable. Any disturbance will lead to shape changes, leading to dissipation. So, if at all, a new balance of centrifugal and bending forces is reached, the free energy will be reduced such that the previous balanced surface shape and rigid body rotation configuration can not be reached anymore. Considering the permanent presence of small disturbances, \eg\ due to thermal or numeric noise, we postulate the cascading mechanism 'disturbance - force balance reconfiguration - dissipation' to eventually lead to $\velocity = 0$ and the classical equilibrium shapes of the Helfrich energy, characterized in  \cite{seifert1997configurations}.

\section{Conclusion}
\label{sec:conclusion}
Using numerical experiments we have explored recently proposed rotational symmetric equilibrium states of rotating fluid deformable surfaces. We have shows that for the considered parameters these states can be reached. However, they are not stable. Instead we found new configurations. They are also characterized by a balance of centrifugal and bending forces but the rotational axis and the axis of symmetry are not aligned. Such rigid body rotations could be considered as three dimensional Killing vector fields. While having a lower energy, also these states are not stable as again any shape change or perturbation of the velocity field will lead to dissipation and destroys the force balance. However, due to strong mesh deformations associated with such rotations the considered numerical approach was not able to resolve the full evolution towards the equilibrium state. We therefore can only postulate that due to thermal or numerical noise the considered states will relax towards the classical equilibrium shapes of \cite{seifert1997configurations} with $\velocity = 0$ and all rotating equilibrium states are only transient configurations. The numerics predict that these transient states can exist for relatively long times. The existence of these states in applications and their potential relevance in evolution remains an open issue.  

\paragraph{Acknowledgement}
  We acknowledge computing resources provided by ZIH at TU Dresden and by JSC at FZ J\"ulich, within projects WIR and PFAMDIS, respectively. 
  This work was supported by the German Research Foundation (DFG) within the Research Unit ``Vector- and Tensor-Valued Surface PDEs'' (FOR 3013).
  Data that support the findings of this study are available from the corresponding author upon reasonable request. We further thank M. Olshanskii for providing the code to compute the rotating equilibrium shapes in the rotational symmetric setting and V. Krause for his support with the surface finite element algorithm for the full model.

\bibliographystyle{alpha}
\bibliography{lib}   

\end{document}